\documentclass[fleqn,10pt]{wlscirep}
\usepackage[utf8]{inputenc}
\usepackage[T1]{fontenc}
\usepackage{chngcntr}
\usepackage{afterpage}
\title{Acoustic Transmission Loss in Hilbert Fractal Metamaterials}

\author[1,*]{Gianni Comandini}
\author[2]{Morvan Ouisse}
\author[1,3]{Valeska P. Ting}
\author[1]{Fabrizio Scarpa}
\affil[1]{Bristol Composite Institute (BCI), School of Civil, Aerospace and Mechanical Engineering (CAME), University of Bristol, Bristol, United Kingdom}
\affil[2]{SUPMICROTECH, University of Franche-Comté, CNRS Institut FEMTO-ST, FR-25000 Besançon, France}
\affil[3]{Research School of Chemistry, The Australian National University, Canberra ACT 2602, Australia}
\affil[*]{gianni.comandini@bristol.ac.uk}


\begin{abstract}
Acoustic metamaterials are increasingly being considered as a viable technology for sound insulation. Fractal patterns constitute a potentially groundbreaking architecture for acoustic metamaterials. We describe in this work the behaviour of the transmission loss of Hilbert fractal metamaterials used for sound control purposes. The transmission loss of 3D printed metamaterials with Hilbert fractal patterns related to configurations from the zeroth to the fourth order is investigated here using impedance tube tests and Finite Element models. We evaluate, in particular, the impact of the equivalent porosity and the relative size of the cavity of the fractal pattern versus the overall dimensions of the metamaterial unit. We also provide an analytical formulation that relates the acoustic cavity resonances in the fractal patterns and the frequencies associated with the maxima of the transmission losses, providing opportunities to tune the sound insulation properties through control of the fractal architecture. 
\end{abstract}
\begin{document}

\flushbottom
\maketitle
%
%
\thispagestyle{empty}

\section*{Introduction}

\begin{figure}[h]
\centering
\includegraphics[width=0.6\textheight]{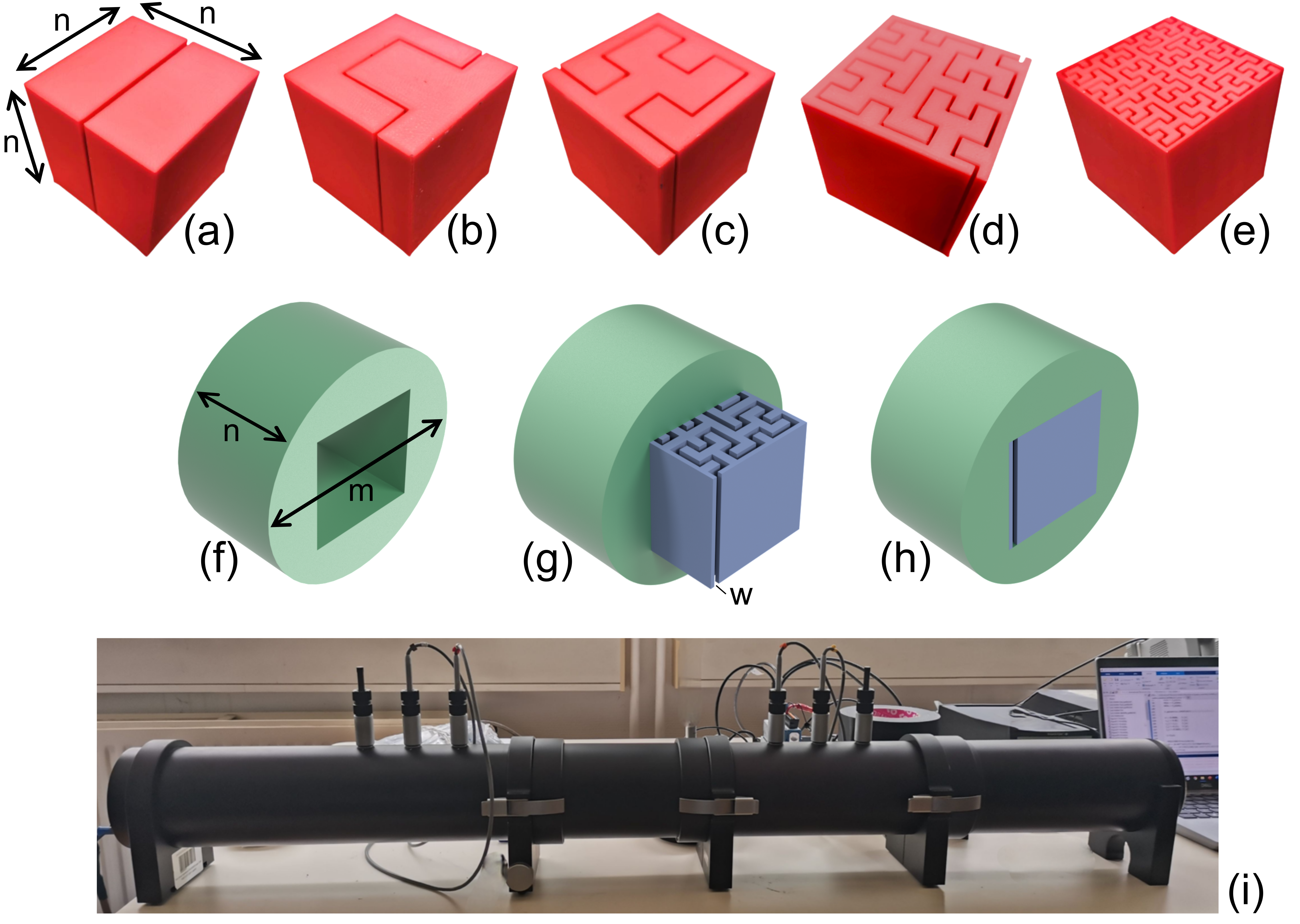}
\caption{The Hilbert fractal acoustic metamaterials and the impedance tube used in this work. (a) The $0^{th}$ order with underlined \textit{\sffamily n}, the side dimension of the cube - 50.8 $mm$, in this case. (b) $1^{st}$, (c) $2^{nd}$, (d) $3^{rd}$, and (e) $4^{th}$ orders of the Hilbert fractal. (f) Sample holder used for the samples inside the test room of the impedance tube. The holder has an external diameter \textit{\sffamily m} of 100 $mm$ and thickness \textit{\sffamily n} of 50.8 $mm$. (g) Example of the cubic specimen (d), having been partially inserted into the holder (f), showing the gap width \textit{\sffamily w}. (h) Holder and sample completely assembled. (i) The impedance tube.}
\label{fig:1}
\end{figure}

Excessive noise can significantly impact human health and lead to serious pathology\cite{zhou2012environmental,basner2014auditory}. Traditional noise transmission mitigation strategies\cite{liu2022definition,tao2021recent} have followed mass law principles\cite{adkins1963transformer}, with a resulting drawback in terms of high mass and thickness of the materials used\cite{may1980optimum}. Mass-spring-mass models exploit other mechanisms, with the development of designs based on one or multiple coupled degrees of freedom and combinations of materials\cite{de2020metamaterial}. Spring-mass models are typically used in acoustic architecture to separate environments with rational use of the space available. Other more recent but consolidated technologies exploit the use of porous materials as perforated\cite{starkey2019experimental,d2012broadband,magliacano2020computation} and/or microperforated panels\cite{bravo2012sound}, or through thermo-viscous acoustic energy dissipation in the internal porous skeletons of foams with various internal geometry and materials compositions\cite{kuhl1932absorption,kuczmarski2011acoustic,ba2017soft,caniato2022sustainable,jia2020highly}. Thermo-viscous energy dissipation technologies also theoretically allow acoustic dampening of any frequency, with the only limitation represented by the available room for the porous liners.
Because of space constraint limitations, porous materials are generally adopted to mitigate noise within the middle and high-frequency ranges. Another technique to obtain a desired transmission loss (TL) for a predetermined interval of frequencies is by using Helmholtz resonators\cite{romero2016perfect,huang2021sound,long2018reconfigurable,lan2017manipulation}. However, resonators are only efficient for narrow sets of frequency ranges and may involve the use of considerable space for the spring part of the Helmholtz resonator for low frequencies applications\cite{tang1973theory,abbad2019numerical}.
More recent techniques for noise management are promising because of their lightweight characteristics and their use of different mechanisms to achieve sound absorption and transmission loss\cite{cummer2016controlling,zhu2016implementation,kadic20193d,assouar2018acoustic,zhao2023scalable,jia2020highly,jiang2015ultralight,qiu2012biomimetic,si2014ultralight}. Metamaterials for noise control are moving from conceptual design\cite{xue2022topological,yin2018band,beli2019wave} to practical implementation\cite{morandi2016standardised,xie2014wavefront,singh2022hybrid,gupta2023metamaterial, tang2017hybrid,xie2018acoustic,song2014sound,d2019low}.
Fractal geometries have been used to design electromagnetic and mechanical metamaterials\cite{park2017unusually,jiang20183d,lv2014origami,gatt2015hierarchical}. Moreover, fractal mechanical metamaterials possess  strength, shape stability under large deformations, and significant damage tolerance\cite{zhang2021harnessing}. The Hilbert lattice\cite{hilbert1935stetige} is one of the most heavily explored fractals in metamaterial design, especially for electronics/electromagnetic applications. Fractal antennas have shown enhanced performance compared to traditional designs since they can operate at multiple frequencies without load and are more compact than those featuring other geometries\cite{thekkekara2017bioinspired,maragkou2014fractal, fan2014fractal,elwi2019printed}.
Recent studies in the field of acoustics have improved the understanding of the Hilbert space-filling curve pattern in metamaterials, providing numerous benefits ranging from high porosity, multiple resonances, multi-modes, and sub-wavelength scales\cite{comandini2022sound,song2016broadband}.
This work describes, for the first time, a numerical and experimental evaluation of the acoustic transmission loss capability of Hilbert fractal metamaterials. The fractal topologies covered here range from the zeroth (Fig.\ref{fig:1}$a$) to the fourth order (Fig.\ref{fig:1}$e$). We have used additive manufacturing to generate samples of fractal metamaterials made using polylactic acid (PLA) because of its recyclability and good printability. One of the design features of the fractal specimens is the presence of inlets and outlets for the impinging and exiting acoustic waves (Fig.\ref{fig:1}$a-e$). The geometry of the fractal patterns is here defined in nondimensional terms, using the lateral dimension of the metamaterial sample \textit{\sffamily n} (50.8 $mm$ in our case), the gap width \textit{\sffamily w} (see Fig.\ref{fig:1}g), and the fractal order $\xi$. The parameter $\eta=\textit{\sffamily w}/\textit{\sffamily n}$ represents the gap width and the porosity - the latter is  $\varphi=\eta(2^{\xi})$ by inspection. An impedance tube with four microphones was used to obtain the experimental measurements of the TL (Fig.\ref{fig:1}$i$). Numerical Finite Element simulations using COMSOL Multiphysics\cite{user2020guide} featuring different models representing the air inside the cavities of the fractal metamaterials, and analytical models were also performed. The models were used here to benchmark the results and understand some of the physical mechanisms underpinning the amplitude and frequency of the TL peaks.


\begin{figure}
\centering
\includegraphics[width=0.60\textheight]{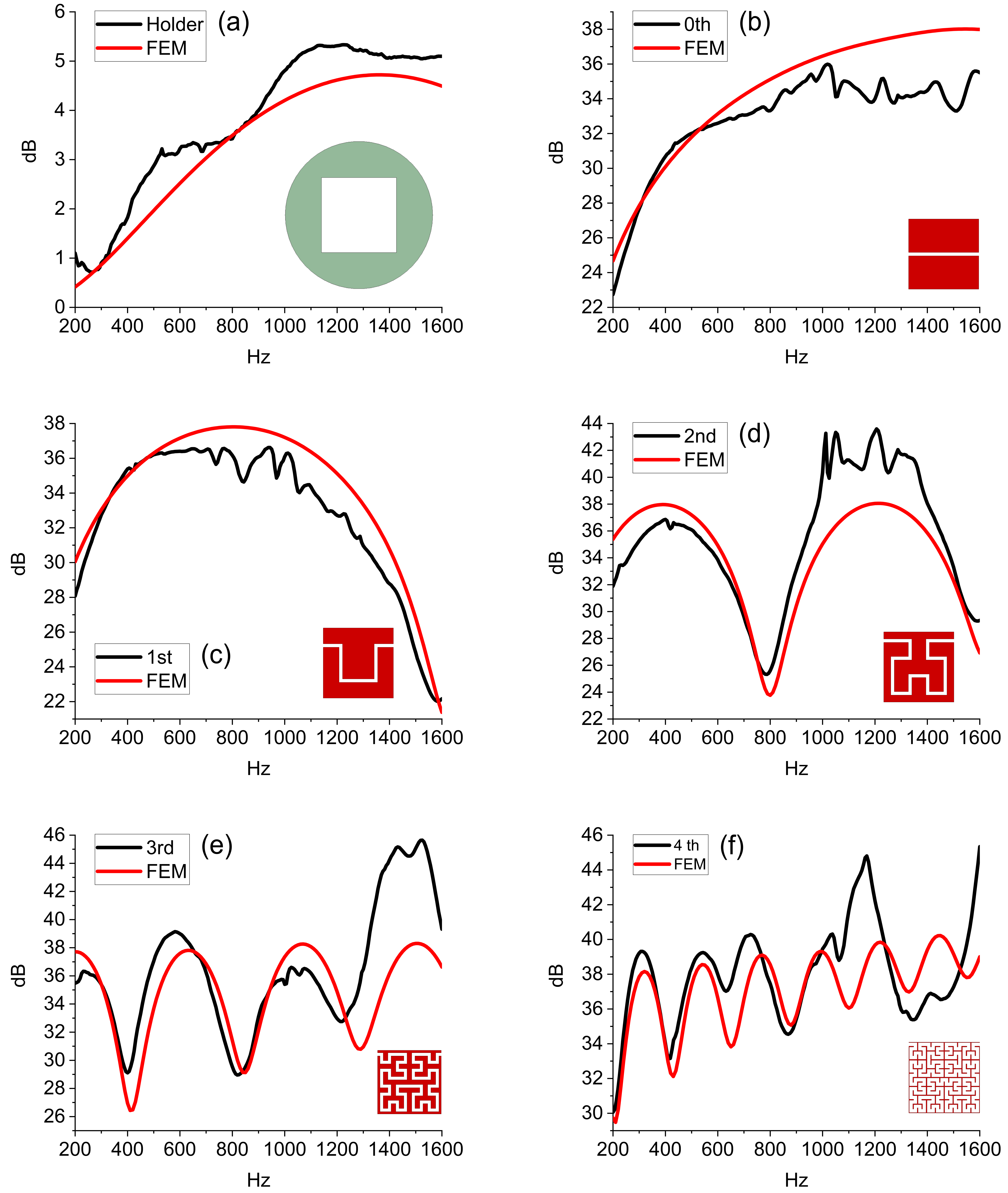}
\caption{Comparison of the transmission loss between experimental results (black) and FEM (red) for the various Hilbert fractal orders. All the measurements and simulations were made taking into account the cylindrical holder of Fig.\ref{fig:1}$f$. (a) Only the holder (Fig.\ref{fig:1}$f$). (b) Zeroth (Fig.\ref{fig:1}$a$). (c) First (Fig.\ref{fig:1}$b$). (d) Second (Fig.\ref{fig:1}$c$). (e) Third (Fig.\ref{fig:1}$d$). (f) Fourth (Fig.\ref{fig:1}$e$) orders.}
\label{fig:2}
\end{figure} 
\noindent 

\section*{Results and Discussion}
We performed a comparison between experimental and finite element-simulated transmission losses for the Hilbert fractal metamaterials with a gap width of 1.0 $mm$ (i.e., corresponding to a value of $\eta$ equal to 2 \%). The size of the gap is the same on both sides of the sample. The finite element results here are related to narrow-acoustics models of the cavity  \cite{user2020guide} inside the fractal metamaterials. There is a good agreement between experiments and simulations, as can be seen in Fig.\ref{fig:2}. A description of the model, and a comparison between lossless, viscous and experimental results related to configurations from the $0^{th}$ to the $4^{st}$ fractal order are presented in the supplementary information. Both viscous-based and lossless formulations for the fluid are able to capture the overall transmission loss properties inside the fractal cavities, and the narrow-acoustics simplification is adequate to follow the experimental results. The lossless model however provides large dips of transmission loss at frequencies slightly larger than those predicted by the viscous models, and provided by the experimental results. It is interesting to note the behaviour of the TL for the various configurations. The order zero (i.e., the topology with the cavity extending to the full length of the cubic sample - Fig.\ref{fig:1}$a$). has a peak of the transmission loss equal to 37 $dB$ at 1546 $Hz$ (Fig.\ref{fig:2}$b$). A peak of 37 $dB$ is however observed at 785 $Hz$ in the case of the first fractal order  (Fig.\ref{fig:2}$c$). The second Hilbert fractal order possesses two peaks at 390 $Hz$ and 1212 $Hz$, both with TL values of 37 $dB$ and 38 $dB$, respectively (Fig.\ref{fig:2}$d$). The third order shows the presence of four transmission loss peaks at 195 $Hz$, 611 $Hz$, 1033 $Hz$, and 1455 $Hz$, all with a TL value of 38 $dB$ (Fig.\ref{fig:2}$e$). Finally, the fourth fractal order exhibits seven TL peaks at 310 $Hz$, 526 $Hz$, 743 $Hz$, 961 $Hz$, 1180 $Hz$, 1399 $Hz$, and 1618 $Hz$, with TL values of 38 $dB$, 38 $dB$, 39 $dB$, 39 $dB$, 40 $dB$, 40 $dB$, and 41 $dB$, respectively (Fig.\ref{fig:2}$f$).  We also include the results related to the transmission losses of the empty holder (Fig.\ref{fig:1}$f$); as expected, the TL values here were the lowest amongst all the configurations considered (Fig.\ref{fig:2}$a$). In summary, both experiments and simulations show the presence of multiple peaks and minima of the transmission loss; the number of those peaks and dips depends on the fractal order of the pattern of the metamaterial considered. The maxima of the transmission loss all show a nearly constant magnitude, with their position dependent on the length of the fractal and, consequently, on a higher Hausdorff dimension\cite{he2009hilbert}. Despite the complexity of the acoustic behaviour of these fractal-shaped metamaterials, it is possible to identify some general trends in their TL response. Increased fractal orders feature larger numbers of TL peaks with a very similar magnitude (see Table~\ref{table2} and Fig.\ref{fig:3}$c$)). This behaviour can be explained by observing that the performance of the TL magnitude of the metamaterial is mainly dependent upon the geometry of the opening slot, rather than on the fractal order itself. 

\begin{table}[ht]
\centering
\begin{tabular}{|c|c|c|c|}
\hline
 Gap Width $\eta$ & Fractal Order $\xi$ & Porosity $\varphi$ & TL [dB] \\
 \hline
 $2\%$ & $0^{th}$, $1^{st}$, $2^{nd}$, $3^{rd}$, $4^{th}$  & $2\%, 4\%, 8\%, 16\%, 31\% $ & $37$ \\
\hline
 $4\%$ & $0^{th}$, $1^{st}$, $2^{nd}$, $3^{rd}$, $4^{th}$ & $4\%, 8\%, 16\%, 31\%, 63\% $ & $31$ \\
\hline
 $6\%$ & $0^{th}$, $1^{st}$, $2^{nd}$, $3^{rd}$, $4^{th}$ & $6\%, 12\%, 24\%, 47\%, 94\% $ & $27$ \\
\hline
 $8\%$ & $0^{th}$, $1^{st}$, $2^{nd}$, $3^{rd}$ & $8\%, 16\%, 31\%, 63\% $ & $25$ \\
\hline
 $10\%$ & $0^{th}$, $1^{st}$, $2^{nd}$, $3^{rd}$ & $10\%, 20\%, 39\%, 79\% $ & $23$ \\
\hline
 $12\%$ & $0^{th}$, $1^{st}$, $2^{nd}$, $3^{rd}$ & $12\%, 24\%, 47\%, 94\% $ & $21$ \\
\hline
$14\%$ & $0^{th}$, $1^{st}$, $2^{nd}$ & $14\%, 28\%, 55\% $ & $19$ \\
\hline
$16\%$ & $0^{th}$, $1^{st}$, $2^{nd}$ & $16\%, 31\%, 63\% $ & $18$ \\
\hline
$18\%$ & $0^{th}$, $1^{st}$, $2^{nd}$ & $18\%, 35\%, 71\% $ & $17$ \\
\hline
$20\%$ & $0^{th}$, $1^{st}$, $2^{nd}$ & $20\%, 39\%, 79\% $ & $16$ \\
\hline
 \end{tabular}
 \caption{\label{tab:2} Normalized gap, fractal orders, porosity, and TL values for the classes of fractal metamaterials evaluated in this work. The leftmost column shows the normalized gap values, while the second column displays the corresponding fractal orders, which are proportional to the gap width. The third column contains the equivalent porosity of each fractal order in terms of relative gap width. 
 The fourth column shows the values of the peaks of the transmission loss calculated via FEM and related to the first cavity mode of the fractals for a given gap width.}
 \label{table1}
 \end{table}

\begin{table}[ht]
\centering
\begin{tabular}{|c|c|c|c|c|}
\hline
Fractal Order $\xi$ & Mode Number & FEM Mean [Hz] & Normalised Standard Deviation  & Eq. \ref{equation1} \\
\hline
$0^{th}$ & $1^{st}$ & $1472.2$ & $2.8\%$ & $1688$ \\
\hline
$1^{st}$ & $1^{st}$ & 811.2 & 0.7\% & 844 \\
\hline
$2^{nd}$ & $1^{st}$ & $426.6$ & $1.1\%$ & $422$ \\
\hline
$2^{nd}$ & $2^{nd}$ & $1295.3$ & $2.7\%$ & $1266$ \\
\hline
$3^{rd}$ & $1^{st}$ & $224.8$ & $1.2\%$ & $211$ \\
\hline
$3^{rd}$ & $2^{nd}$ & $685.3$ & $3.1\%$ & $63$3 \\
\hline
$3^{rd}$ & $3^{rd}$ & $1148$ & $4.8\%$ & $1055$ \\
\hline
$3^{rd}$ & $4^{th}$ & $1602.5$ & $6.2\%$ & $1477$ \\
\hline
$4^{th}$ & $1^{st}$ & $328$ & $1.0\%$ & $316$ \\
\hline
$4^{th}$ & $2^{nd}$ & $553.7$ & $1.6\%$ & $527$ \\
\hline
$4^{th}$ & $3^{rd}$ & $781$ & $2.2\%$ & $73$8 \\
\hline
$4^{th}$ & $4^{th}$ & $1008$ & $2.7\%$ & $949$ \\
\hline
$4^{th}$ & $5^{th}$ & $1235.3$ & $3.2\%$ & $1160$ \\
\hline
$4^{th}$ & $6^{th}$ & $1462.$7 & $3.7\%$ & $1371$ \\
\hline
$4^{th}$ & $7^{th}$ & $1689.3$ & $4.1\%$ & $1582$ \\
\hline
\end{tabular}
\caption{\label{tab:1} Cavity resonances predicted via Finite Elements (narrow viscous model) and the analytical formulation for the various fractal orders. The Finite Element results related to each mode are calculated by averaging the values of those modes over the different gap widths considered in this work. The standard deviations of those frequencies are normalised against the corresponding average value. The last column refers to the analytical resonance frequency related to an open-closed cavity.} 
\label{table2}
\end{table}

\afterpage{
\begin{figure}[ht]
\centering
\includegraphics[width=0.65\textheight]{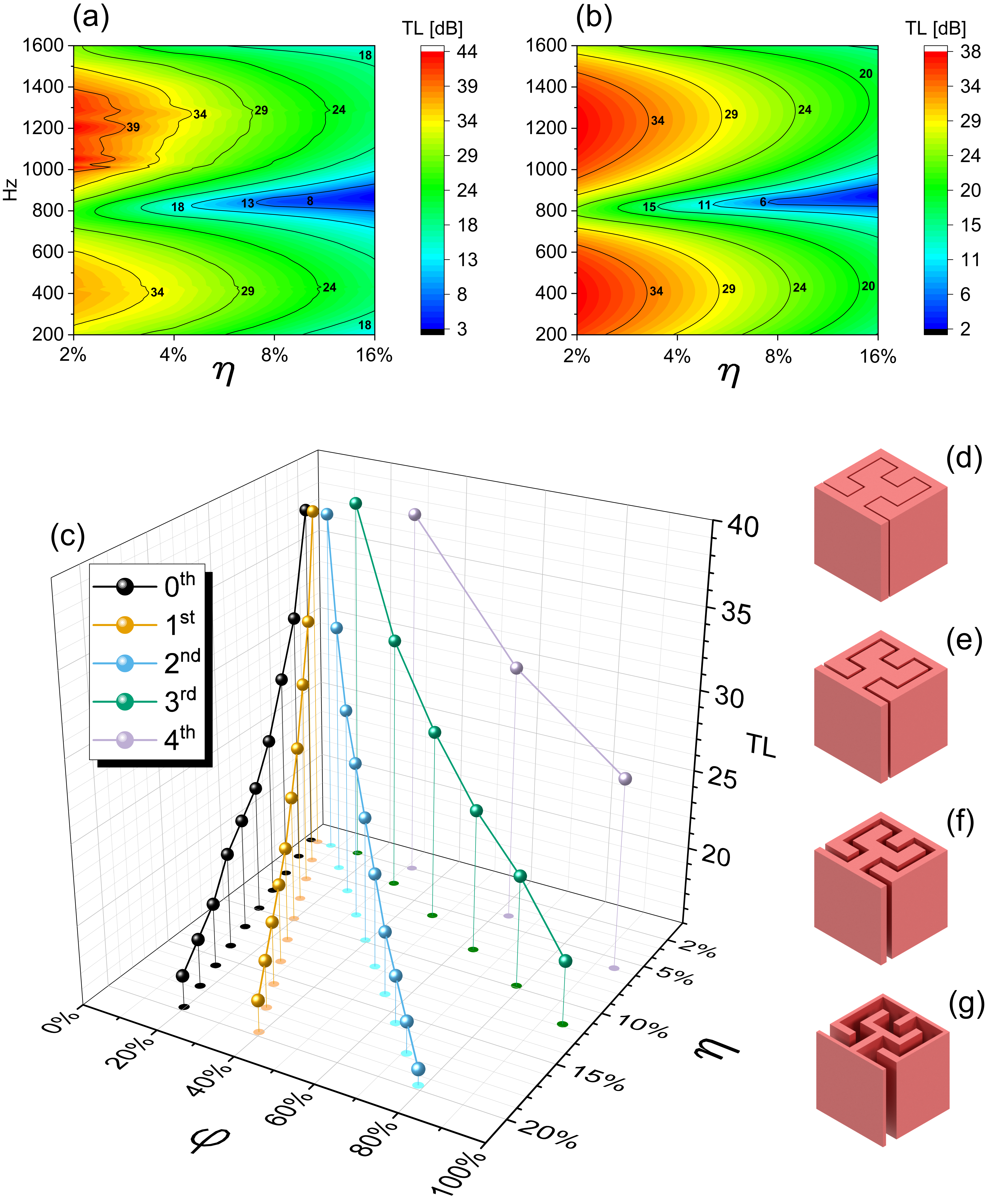}
\caption{Experimental (a) and FE (b) contour plots of the TL  as a function of the frequencies and normalised gap widths. (c) Variation of the transmission loss peaks calculated via FEM for the fractal orders and their porosity when the normalised gap changes. Example of the $2^{nd}$ order Hilbert fractal with a gap width of (d) $1\ mm$, (e) $2\ mm$, (f) $4\ mm$, and (g) $8\ mm$. }
\label{fig:3}
\end{figure}
\clearpage
}

Table~\ref{table1} shows the dependence of the magnitude of the transmission loss peak values from the FEM simulations versus the normalised gap width, the fractal orders and the porosity of the metamaterials. One can notice that the most significant factor affecting the values of the transmission losses is the gap width, regardless of the fractal order or porosity considered (see also Fig.\ref{fig:3}$c$). 
Of particular interest is also the effect of the fractal orders on the frequencies corresponding to the peak of the transmission losses, which are also known to affect the acoustic absorption properties of Hilbert fractal metamaterials\cite{comandini2022sound}. The results suggest that while the order of the fractal pattern can impact both peak TL magnitudes and related frequencies, it is not the dominant factor to control those frequencies in configurations with varying normalised gap widths. Furthermore, the results indicate that variations of the porosity have a limited effect on the transmission losses (Fig.\ref{fig:3}$c$ and Table~\ref{table1}).

 The results of the experimental and numerical transmission losses show that the frequencies corresponding to the peak of the TL can be approximated using equation (Eq.\ref{equation1}), which is also adopted for the design of musical (wind) instruments\cite{ruiz2017hearing,ayers1985conical}. Formula (Eq.~\ref{equation1}) is related to open-closed one-dimensional resonating cavities and considers even harmonics only (i.e., $r = m+1$ with $m\in \mathbb{N}$) \cite{ruiz2017hearing}. The resonant frequency for an open-closed cavity representing the fractal path is equal to:

\begin{equation}
f_{n} = \frac{rc}{4n(2^{\xi})}
\label{equation1}
\end{equation}

\noindent 

In Eq.~\ref{equation1}, $c$ is the speed of sound in air, $343\ m/s$ and $L=n(2^{\xi})$ is the length of the fractal cavity inside the metamaterial sample. At the frequencies corresponding to the various cavity modes of the fractal patterns, the peak of the TL is always close to a value of 37 $dB$ if the opening percentage on both sides of the metamaterial is $2\%$, no matter the fractal order considered (Table~\ref{table1}). This simplification is valid, especially for the case of the first peaks related to the first acoustic modes in the lower frequency range (Table~\ref{table1}). 
Table~\ref{table2} summarises the numerical (narrow-acoustic approximation\cite{user2020guide}) and analytical results related to the TL for different gap widths and fractal orders. The vibrational modes corresponding to the TL peaks are also indicated. The standard deviation of the distribution of frequencies associated with transmission loss peaks for the different gap widths is normalised against the value of 1688 Hz, which corresponds to the average frequency of the TL peak related to the $0^{th}$ order when the gap width ($\eta$) is varied from 2\% to 20\% (Fig.\ref{fig:3}d-g). Notably, the frequency corresponding to the TL peaks predicted by the analytical open-closed resonator formula is also close to the numerical FE one (Table~ \ref{table2}, fifth column). The number of vibrational modes within the frequency range investigated here increases with the increasing fractal order, and the frequencies related to the TL peaks shift to higher values. The magnitudes of TL peaks are again almost constant, with only slight variations attributed to the change in the length of the fractal pattern inside the metamaterial. These findings imply that by increasing the fractal order, it is possible to achieve more resonant peaks, resulting in more effective broadband transmission loss, and also sound absorption \cite{comandini2022sound}. Moreover, both numerical and analytical modelling approaches provide a good approximation of the actual behaviour of these fractal acoustic metamaterials (Fig.\ref{fig:2}, and Fig.\ref{fig:3}$a-b$).

\noindent

\section*{Conclusion}

This work has shown the relationship between the fractal order and the transmission loss in acoustic metamaterials with Hilbert fractal patterns. The equivalent porosity and gap widths play a crucial role in determining the values of the transmission loss (Fig. \ref{fig:3}$c$), with the gap width being the most significant factor impacting the magnitudes of the TL (Fig.\ref{fig:3}$a-b$ and Table~\ref{table1}). Our experimental and numerical results show that a decrease in the transmission loss is generated by the increase of the gap width, with a consequent reduction of the performance of the acoustic metamaterial. 
We have also investigated the influence of fractal order on the transmission loss. Our simulations show that the fractal order plays a role in determining the number of TL peaks, but not on their magnitude. 
Finally, the frequencies corresponding to the peaks of the transmission losses can be well approximated and predicted by considering the fractal patterns as open-closed one-dimensional cavities, and by calculating the corresponding resonance frequencies. The formula \ref{equation1} makes it possible at the design stage to relate those frequencies to the fractal order and nondimensional gap width of the fractal acoustic metamaterial. 

\section*{Methods}

All the samples were printed using the Raise3D Pro3 Plus 3D Printer. The machine was set up with a 0.1 $mm$ layer height, 100\% infill density, and an extrusion nozzle of 0.4 $mm$. 3DJake recycled PLA was used as the filament. The transmission loss have been measured following the ASTM E2611-09 standard \cite{ASTM:E2611-09}, using a Brüel \& Kjær impedance tube (Fig.\ref{fig:1}$i$) with a four microphones configuration and double load measurements. The gap widths of the metamaterials samples  are 1 $mm$, 2 $mm$, 4 $mm$ and 8 $mm$, corresponding to 2\% (Fig.\ref{fig:3}$d$), 4\% (Fig.\ref{fig:3}$e$), 8\% (Fig.\ref{fig:3}$f$) and 16\% (Fig.\ref{fig:3}$g$). All the tests were performed by mounting the fractal metamaterial sample in a cylindrical holder (Fig.\ref{fig:1}$f-h$). The FEM simulations were performed using the Comsol Multiphysics 5.6 software\cite{user2020guide} with a 3D model of the impedance tube and the metamaterials specimens \cite{comandini2022sound}. The FEM model does not consider the presence of the PLA material, only the fluid cavity is modelled, with loss-less,  thermoviscous and narrow-viscous domains. The part of the impedance tube model outside the test room was represented with a lossless air fluid, with probes simulating the position of the four microphones. The frequency range investigated in both FEM and experiments was from 200 $Hz$ to 1600 $Hz$.

\section*{Acknowledgements}

G.C. acknowledges the support of UK EPSRC through the EPSRC Centre for Doctoral Training in Advanced Composites for Innovation and Science [EP/L016028/1], and Rafael Heeb, Emanuele De Bono and Martin Gillet for technical assistance. F.S. acknowledges the support of ERC-2020-AdG-NEUROMETA (No. 101020715). VT acknowledges funding from the EPSRC (EP/R01650X/1).

\section*{Author contributions statement}

G.C., M.O, and F.S. conceived the experiments,  G.C conducted the experiments, G.C conducted the simulations, G.C., M.O, V.T., and F.S. analysed the results.  All authors reviewed the manuscript. 

\section*{Competing interests}
The authors declare no conflicts of interest exist for the development of this work.

\bibliography{sample}

\end{document}